\documentclass[aps,10.4pt,twocolumn, twoside, nofootinbib]{revtex4}
\usepackage{amsthm}
\usepackage{amsmath,latexsym,amssymb,verbatim,enumerate,graphicx}
\usepackage{color}

\newtheorem{definition}{Definition} 

\newtheorem{thm}[definition]{Theorem}
\newtheorem{corollary}[definition]{Corollary}

\newtheorem*{rep@theorem}{\rep@title}
\newcommand{\newreptheorem}[2]{%
\newenvironment{rep#1}[1]{%
 \def\rep@title{#2 \ref{##1} (restatement)}%
 \begin{rep@theorem}}%
 {\end{rep@theorem}}}
\makeatother

\newreptheorem{thm}{Theorem}
\newreptheorem{lem}{Lemma}

\def\ba#1\ea{\begin{align}#1\end{align}}
\def\ban#1\ean{\begin{align*}#1\end{align*}}

\newcommand{\ot}{\otimes}
\newcommand{\be}{\begin{equation}}
\newcommand{\ee}{\end{equation}}

\def\benum{\begin{enumerate}}
\def\eenum{\end{enumerate}}

\def\squareforqed{\hbox{\rlap{$\sqcap$}$\sqcup$}}
\def\qed{\ifmmode\squareforqed\else{\unskip\nobreak\hfil
\penalty50\hskip1em\null\nobreak\hfil\squareforqed
\parfillskip=0pt\finalhyphendemerits=0\endgraf}\fi}
\def\endenv{\ifmmode\;\else{\unskip\nobreak\hfil
\penalty50\hskip1em\null\nobreak\hfil\;
\parfillskip=0pt\finalhyphendemerits=0\endgraf}\fi}

\newcommand{\bra}[1]{\langle #1|}
\newcommand{\ket}[1]{|#1\rangle}

\newcommand{\tr}{\text{tr}}

\newcommand{\id}{\mathbb{I}}

\newcommand{\<}{\langle}
\renewcommand{\>}{\rangle}

\def\id{{\operatorname{id}}}

\def\be{\begin{equation}}
\def\ee{\end{equation}}
\def\ben{\begin{eqnarray}}
\def\een{\end{eqnarray}}
\def\ot{\otimes}

\def\bei{\begin{itemize}}
\def\eei{\end{itemize}}

\mathchardef\ordinarycolon\mathcode`\:
\mathcode`\:=\string"8000
\def\vcentcolon{\mathrel{\mathop\ordinarycolon}}
\begingroup \catcode`\:=\active
  \lowercase{\endgroup
  \let :\vcentcolon
  }

\newcommand{\nc}{\newcommand}
 \nc{\proj}[1]{|#1\rangle\!\langle #1 |} 
\nc{\avg}[1]{\langle#1\rangle}

\nc{\conv}{\operatorname{conv}}
\nc{\smfrac}[2]{\mbox{$\frac{#1}{#2}$}} \nc{\Tr}{\operatorname{Tr}}
\nc{\ox}{\otimes} \nc{\dg}{\dagger} \nc{\dn}{\downarrow}
\nc{\lmax}{\lambda_{\text{max}}}
\nc{\lmin}{\lambda_{\text{min}}}

\nc{\csupp}{{\operatorname{csupp}}}
\nc{\qsupp}{{\operatorname{qsupp}}} \nc{\var}{\operatorname{var}}
\nc{\rar}{\rightarrow} \nc{\lrar}{\longrightarrow}
\nc{\poly}{\operatorname{poly}}
\nc{\polylog}{\operatorname{polylog}} \nc{\Lip}{\operatorname{Lip}}
\nc{\Om}{\Omega}
\nc{\wt}[1]{\widetilde{#1}}

\def\>{\rangle}
\def\<{\langle}

\nc{\glneq}{{\raisebox{0.6ex}{$>$}  \hspace*{-1.8ex} \raisebox{-0.6ex}{$<$}}}
\nc{\gleq}{{\raisebox{0.6ex}{$\geq$}\hspace*{-1.8ex} \raisebox{-0.6ex}{$\leq$}}}

\nc{\vholder}[1]{\rule{0pt}{#1}}
\nc{\wh}[1]{\widehat{#1}}
\nc{\h}[1]{\widehat{#1}}

\nc{\ob}[1]{#1}

\def\beq{\begin {equation}}
\def\eeq{\end {equation}}

\def\be{\begin{equation}}
\def\ee{\end{equation}}

\nc{\eq}[1]{(\ref{eq:#1})} 
\nc{\eqs}[2]{\eq{#1} and \eq{#2}}

\nc{\eqn}[1]{Eq.~(\ref{eqn:#1})}
\nc{\eqns}[2]{Eqs.~(\ref{eqn:#1}) and (\ref{eqn:#2})}

\nc{\region}{\cS\cW}

\begin{document}

\title{{\Large Entanglement Area Law from Exponential Decay of Correlations }}

\author{Fernando G.S.L. Brand\~ao}
\email{f.brandao@ucl.ac.uk}
\affiliation{Department of Computer Science, University College London}
\affiliation{National Quantum Information Center of Gdansk}

\author{Micha\l{} Horodecki}
 \email{fizmh@ug.edu.pl}
\affiliation{Institute for Theoretical Physics and Astrophysics, University of Gda\'nsk, 80-952 Gda\'nsk, Poland}


\begin{abstract}

Area laws for entanglement in quantum many-body systems give useful information about their low-temperature behaviour and are tightly connected to the possibility of good numerical simulations. An intuition from quantum many-body physics suggests that an area law should hold whenever there is exponential decay of correlations in the system, a property found for instance in non-critical phases of matter. However the existence of quantum data-hiding states -- i.e. states having very small correlations, yet a volume scaling of entanglement -- was believed to be a serious obstruction to such an implication. Here we prove that notwithstanding the phenomena of data hiding, one-dimensional quantum many-body states satisfying exponential decay of correlations always fulfil an area law. To obtain the result we combine several recent advances in quantum information theory, thus showing the usefulness of the field for addressing problems in other areas of physics. 



\end{abstract}

\maketitle

\parskip .75ex



Certain properties of the lowest-energy state, the groundstate, of a quantum many-body Hamiltonian give a lot of information about 
the physics of the model at zero or low temperatures. A well-studied property is the decay of correlations between 
different regions, which can be used to identify the phase of the model \cite{Sac01}. Another property  is the amount of entanglement in the state. It was first analysed in relation to the black hole entropy problem \cite{Bek71, Haw74, BKLS86} and more recently also in the context of quantum many-body physics \cite{ECP10, VLRK03, CC04, PEDC05, Wol06, Has07}. This is an interesting quantity to study not only because of the resource character of entanglement in quantum information science \cite{HHHH08, BBPS96}, but also because it can be used to elucidate aspects of the physics of the system, such as whether it is close to criticality \cite{ECP10, Wol06}. Another motivation comes from the fact that large amounts of entanglement in a quantum state usually renders its classical simulation infeasible. Thus it is interesting to find out in which cases there is only limited entanglement in the system, which many times is known to lead to good numerical methods to simulate it \cite{Has07, VC06, FNW92, Whi92}. 

Is there a relation between these two properties? To answer this question is the main goal of the paper. But before turning to it, let us define more precisely the two notions. 

\vspace{0.3 cm}

\noindent  \textit{Decay of Correlations:} Given a bipartite (mixed) quantum state $\rho_{XY}$, we quantify the correlations between $X$ and $Y$ by 
\begin{eqnarray}
\label{eq:cor}
&&\text{Cor}(X:Y)  \\ &:=& \max_{\Vert M \Vert \leq 1, \Vert N \Vert \leq 1} |\tr((M \otimes N)(\rho_{XY} - \rho_{X} \otimes \rho_{Y}))|.\nonumber
\end{eqnarray}
where $\Vert M \Vert$ is the operator norm of $M$, given by the maximum eigenvalue of $(M^{\cal y}M)^{1/2}$. Such correlation function generalizes two-point correlation functions, widely studied in condensed matter physics, in which both $X$ and $Y$ are composed of a single site.

We say a quantum state $\ket{\psi}_{1, ..., n}$ composed of $n$ qubits defined on a finite dimensional lattice has $\xi$-exponential decay of correlations if
for every two regions $X$ and $Y$ separated by $l$ sites,
\begin{equation} \label{EDC}
\text{Cor}(X:Y) \leq 2^{- l/\xi}.
\end{equation} 
Here $\xi$ is the correlation length of the state. Such a form of exponential decay of correlations is sometimes also termed the exponential clustering property (see e.g. \cite{AHR62, Fre85, NS06}) and is expected to appear in non-critical phases of matter, where there is a notion of correlation length. Indeed it has been proved that groundstates of gapped Hamiltonians always satisfy Eq. (\ref{EDC}), with a $\xi$ of order of the inverse spectral gap of the model \cite{Has04, Has04b, HT06, NS06}. 

\vspace{0.3 cm}

\noindent  \textit{Scaling of Entanglement:} Given a bipartite pure state $\ket{\psi}_{XY}$ we define the entanglement of $X$ with $Y$ as \cite{HHHH08, BBPS96}.
\begin{equation}
E(\ket{\psi}_{XY}) = H(\rho_X) = - \tr(\rho_X \log \rho_X),
\end{equation}
with $\rho_X$ the reduced density matrix of $\ket{\psi}_{XY}$ on the region $X$ and $H(\rho_X)$ its von Neumann entropy. 

Starting with the work of Bekenstein on the entropy of black holes \cite{Bek71, Haw74, BKLS86} and later also
in the context of quantum spin systems \cite{VLRK03, CC04} and quantum harmonic systems \cite{PEDC05, Wol06}, an increasing body  of evidence appeared suggesting that states corresponding to the ground or to low-lying energy eigenstates 
of local models satisfy an \textit{area law} \cite{ECP10}, i.e. the entanglement of a contiguous region is proportional to its boundary, and not 
to its volume (which is the typical behaviour for most quantum states). In a ground-breaking work, Hastings proved that this is indeed the case for one-dimensional systems \cite{Has07}: 1D gapped Hamiltonians with a unique groundstate always obey an area law, which in this case means that the entanglement of any contiguous region is upper bounded by a constant independent of the size of the region.

\begin{figure}[h]
\begin{center}  
\includegraphics[height=5cm,angle=0]{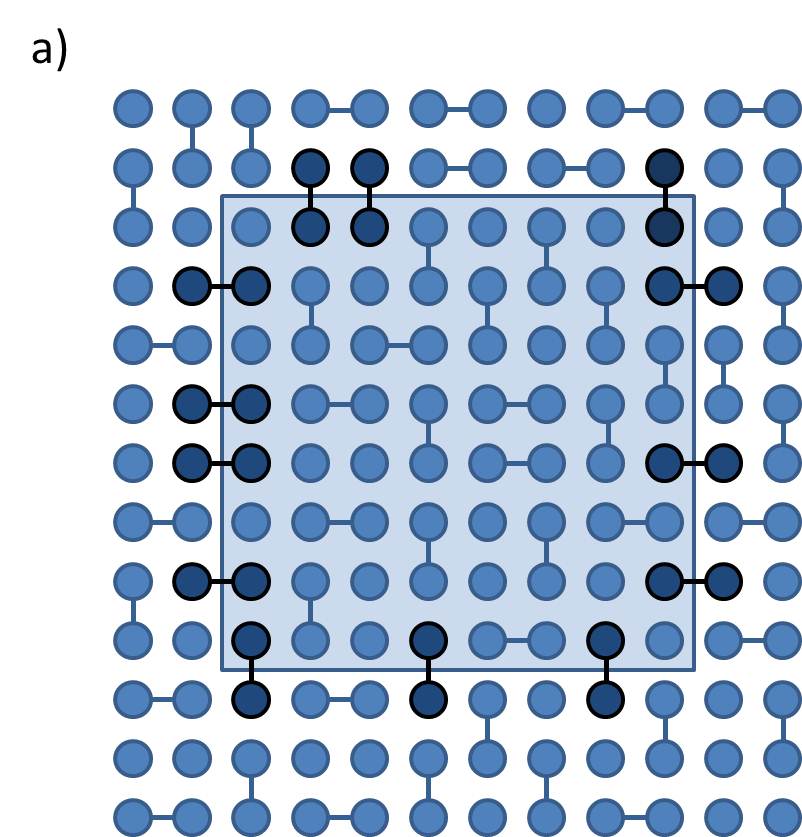}  
\hspace{1cm}
\includegraphics[width=8cm,angle=0]{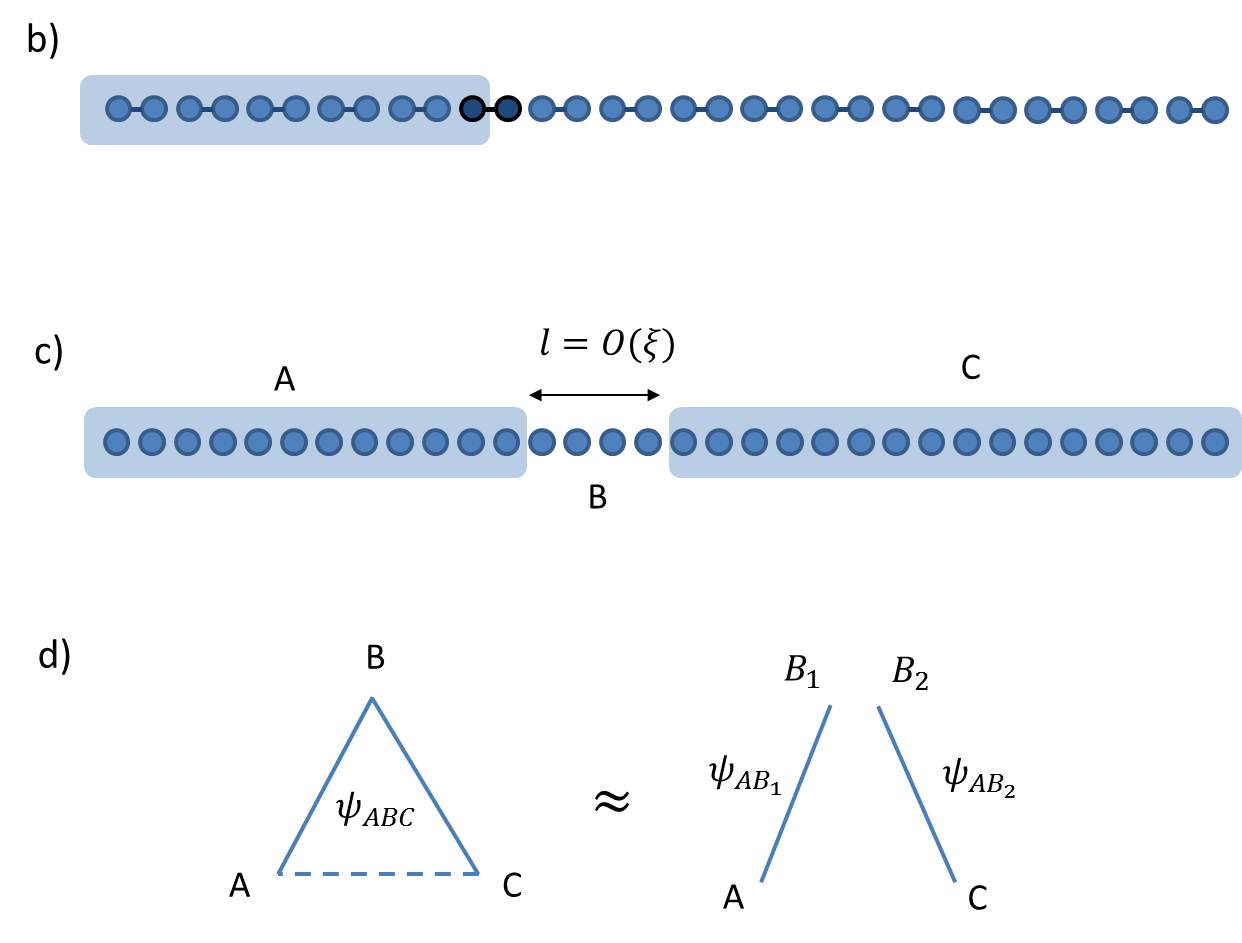}  
\caption{\small  \textit{Exponential decay of correlations intuitively suggests area law}. (a) The intuition is exemplified in a simple manner 
by a state consisting of entangled pairs of neighbouring particles. There the correlations are of fixed length $2$, as only neighbours 
are correlated. The particles connected by an edge are in the pure state $\psi=\frac12(|00\>+|11\>)$, and so only the pairs crossing 
the boundary contribute to the entropy of the region inside the boundary; (b) For 1D states an area law implies that the entropy of an interval 
is constant. Again for a system of entangled pairs, only one pair cut the boundary; (c-d) A general intuitive argument is the following: If the 
distance of two parts $A$ and $C$ is larger than the correlation length, the reduced state $\rho_{AC}$ should be close to a product state: $\rho_{AC}\approx \rho_A\ot \rho_C$, then suggesting that the system $B$ can be divided into subsytems $B_1$ and $B_2$ such that 
the total pure state $\psi_{ABC}$ is close to product state $\psi_{AB_1} \ot \psi_{AB_2}$. However for pure bipartite states, the entropy 
cannot exceed the size of any of subsystems. Therefore $S(A)\leq S(B_1)\approx O(\xi)$, and we would obtain that entropy of any 
interval is constant and proportional to the correlation length $\xi$.
 \label{fig:intuition}}  
\end{center}  
\end{figure}  

\vspace{0.3 cm}

\noindent \textit{Exponential Decay of Correlations versus Area Law:} We can now turn to the question of whether the
two notions introduced above are related. It is a well-known intuition from quantum many-body physics that 
exponential decay of correlations suggests the entanglement of the state should satisfy an area law \cite{VC06}. Indeed, 
consider a quantum state $\ket{\psi}_{ABC}$ as in Fig \ref{fig:intuition}c), with $B$ the boundary region between $A$ and $C$. 
If there is exponential decay of correlations in $\ket{\psi}$ and the separation between $A$ and $C$ is of order 
of the correlation length of the state, $A$ will have almost no correlations with $C$, and one would expect the 
entanglement of $A$ with $BC$ to be only due to correlations with the region $B$, thus obeying an area law.

\begin{figure}[h]
\begin{center}  
\includegraphics[width=8cm,angle=0]{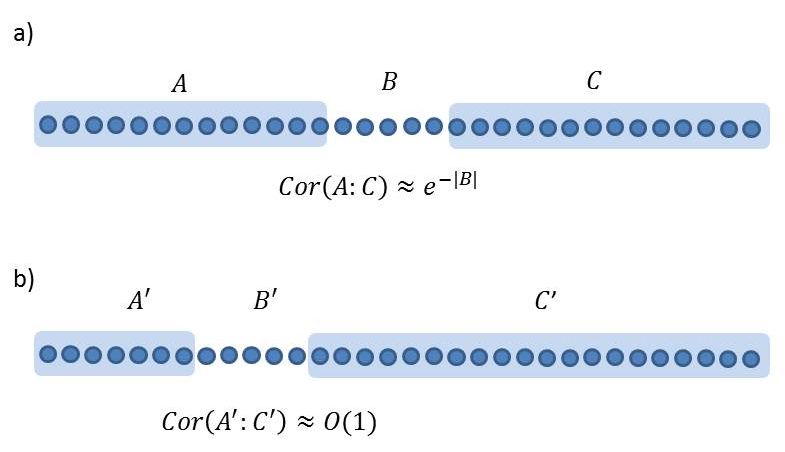}  
\caption{\small \textit{Data hiding as an obstruction}. (a) In quantum information theory there is the phenomenon of data hiding: 
one can find bipartite states $\rho_{AC}$ which can be very well discriminated e.g. from the maximally mixed state if one has access to both subsystems $A$ and $C$, but are indistinguishable from the maximally mixed state by parties that can only make local measurements 
on the subsystems. Such states can be obtained by picking a random pure tripartite state $\ket{\psi}_{ABC}$ 
of $n$ qubits, with $B$ having a small fraction of the total number of qubits. These states have decaying 
correlations (for the above range of sizes of $B$), but their subsystems  $A$ and $C$ are almost maximally mixed, 
hence following a volume law. This shows the intuitive argument of Fig. \ref{fig:intuition} is flawed: the state $\rho_{AC}$ of Fig. \ref{fig:intuition} can be far away from any product state (in a trace norm or fidelity, as needed to carry over the argument outlined in Fig. 1), as the case for data hiding states; (b) However, random states have strong correlations in a different partition. If we divide it into subsystems $A'B'C'$ 
such that the number of qubits of $A'B'$ is smaller than that of $C'$, then $\rho_{A'B'}$ 
is very close to the maximally mixed state $\tau_{A'}\otimes\tau_{B'}$ in trace distance, and by the decoupling argument \cite{Uhlmann76,SchumacherW-decoupling} $C'$ can be divided into $C'_1 C'_2$ such that $\psi_{A'C_1'}$ is maximally entangled. Hence correlations between $A'$ and $C'$ are of order of $1$.
\label{fig:hiding}}  
\end{center}  
\end{figure}

Perhaps surprisingly, the argument presented above is \textit{flawed}. This is because there are quantum states 
for which $A$ and $C$ have almost no correlations, yet $A$ has very large entropy. This is not only a pathological 
case among quantum states, but it is actually the general rule: The overwhelming majority of quantum states will have 
such peculiar type of correlations, whenever the regions $A$ and $C$ have approximately equal sizes \cite{HLW06} (see 
Fig. \ref{fig:hiding}). 
Quantum states with this property are termed quantum \textit{data hiding} states, due to their use in hiding correlations 
(and information) from local measurements \cite{DLT02}. This peculiar kind of quantum correlations has been studied in quantum information theory with an eye on cryptographic applications. Interestingly here they appear as an obstacle to understanding correlations in quantum many-body states. 

Quantum data-hiding states, and the related quantum expander states \cite{Has07b, Has07c}, 
have been largely thought of been an obstruction for obtaining an area law for 
entanglement entropy from exponential decay of correlations (see e.g. \cite{VC06, Has07, Has07b, Has07c, WFHC08, Mas09, AALV10, ALV11, Osb11}). Our main result shows that such a no-go implication, at least for states defined on a 1-dimensional lattice, is in fact false:
\def\entropybound{

Let $\ket{\psi}_{1, ..., n}$ be a state defined on a line  with $\xi$-exponential decay of correlations. Then for any connected region $X \subset [n]$,
\begin{equation} \label{entropybound}
H(X) \leq c \exp\left( c' \log(\xi)\xi \right),
\end{equation}
with $c, c' > 0$ universal constants.}

\begin{thm} \label{maintheorempure}
\entropybound
\end{thm}

The result can be considered as an strengthening of the result of \cite{Has07}, which proved an area law for 1D states with similar parameters under the assumption that the state is a groundstate of a gapped model. 

An interesting class of models to which we can apply the result are (disordered) Hamiltonians exhibiting \textit{many-body localization} (or a mobility gap) \cite{Has10, HSS12}. It has been recently established that their groundstates satisfy exponential decay for correlations \cite{Has10, HSS12}, yet an area law for one-dimensional systems was not known before, since the models are not gapped in general. A direct consequence of Theorem \ref{maintheorempure} is that groundstates of one-dimensional models with a mobility gap \cite{Has10} (or having many-body localization in the sense of \cite{HSS12}) always fulfil an area law; in particular using the result of Ref. \cite{HSS12} that the XY model with random coefficients exhibits many-body localization, we find that its groundstate satisfies an area law.

We note the exponential dependence of the entropy bound with the correlations length. For groundstates of gapped Hamiltonians one has recently sharpened the area law of \cite{Has07} to obtain a linear dependence of entanglement with the spectral gap \cite{AKLV12}. It is an interesting question whether a similar scaling can be obtained merely from the assumption of exponential decay of correlations. Another important open question is whether a similar result can be established in higher dimensions. For $d \geq 2$ an area law is not known even assuming the state is the groundstate of a gapped local Hamiltonian.

\vspace{0.3 cm}

\noindent \textit{Implications and Applications:} One consequence of our result is an approximation of a state $\ket{\psi}_{1, ..., n}$ with exponential decay of correlations in terms of a matrix product state \cite{FNW92, VC06} of small bond dimension. A matrix product representation of the state $\ket{\psi}_{1,...,n}$ is given by 
\begin{equation}
\ket{\psi}_{1,...,n} = \sum_{i_1=1}^d ... \sum_{i_n=1}^d \tr(A^{[1]}_{i_1}...A^{[n]}_{i_n}) \ket{i_1, ..., i_n},
\end{equation}
with $D \times D$ matrices $A^{[j]}$, $j \in [n]$. The parameter $D$ is termed bond dimension and measures the complexity of the matrix product representation. When $D = \poly(n)$ the quantum state $\ket{\psi}_{1,...,n}$ admits an efficient classical description in terms of its matrix product representation, with only polynomially many parameters and in which expectation values of local observables can be calculated efficiently. We call such states themselves matrix product states (MPS).

\begin{corollary} \label{MPS}
Let $\ket{\psi}_{1, ..., n}$ be a state defined on a line with $\xi$-exponential decay of correlations. For every $\delta > 0$ and integer $k$, there is a matrix product state $\ket{\phi_D}$ of bond dimension $D = \poly(k, 1/\delta)$ such that for every contiguous region $R$ of less than $k$ sites, 
\begin{equation}
\Vert \tr_{\backslash R} (\ket{\psi}\bra{\psi})  - \tr_{\backslash R} (\ket{\phi_D}\bra{\phi_D})  \Vert_1 \leq \delta.
\end{equation}
with $\tr_{\backslash R}$ the partial trace over the complement of $R$.
\end{corollary}

We note that strictly speaking the corollary above does not follow directly from Theorem \ref{entropybound}, but rather from a strengthened version given in Theorem \ref{entropyboundsingle} in the methods section (see \cite{BH12} for details).

Thus one-dimensional pure quantum states with exponential decay of correlations have a very simple structure, admitting a classical efficient parametrization. In fact the most successful numerical method presently known for computing low-energy properties of one-dimensional models, the density matrix renormalization group (DMRG) \cite{Whi92}, is a variational method over the class of MPS. Corollary \ref{MPS} shows that one should expect DMRG to work well whenever the model is such that its groundstate has rapidly decaying correlations. 

A second consequence of Theorem I concerns the central question in quantum information science of understanding which properties are behind the apparent superiority of quantum computation over classical computation. A fruitful approach in this direction is to find conditions under which quantum circuits have an efficient classical simulation (see e.g. \cite{Got98, JL02, Val02, dVT02, Vid03, MS08, Nest11}), hence finding properties a quantum circuit \textit{ought} to have if it is supposed to give a superpolynomial speed-up.

An interesting result in this direction is the following \cite{Vid03}: Unless a quantum computation in the circuit model involves states that violate an area law, with the entropy of a certain subregion being bigger than the logarithmic of the number of qubits, it can be simulated classically in polynomial time. A direct consequence of this result and Theorem \ref{maintheorempure} is the following:

\begin{corollary}
Consider a family of quantum circuits $V = V_{k}...V_2V_1$ acting on $n$ qubits arranged in a ring and composed of two qubit gates $V_k$. Let $\ket{\psi_t} := V_{t}...V_2V_1 \ket{0^n}$ be the state after the $t$-th gate has been applied. Then if there is a constant $\xi$ independent of $n$ such that, for all $n$ and $t \in [n]$, $\ket{\psi_t}$ has $\xi$-exponential decay of correlations, one can classically simulate the evolution of the quantum circuit in $\poly(n, k)$ time.
\end{corollary}

The corollary says that one must have at least algebraically decaying correlations in a quantum circuit if it is supposed to solve a classically hard problem more efficiently. Interestingly such kind of long range correlations are usually associated to critical phases of matter. From a quantum information perspective, the result gives a limitation to the possibility of hiding information in 1D quantum circuits: If correlations are hidden in all partitions at most times, the computation can be simulated efficiently classically. Indeed if that is the case, then the state of the quantum computation at most time steps will have an efficient matrix product representation and we can use the results of \cite{Vid03} to give an efficient classical simulation for it. 

\vspace{0.3 cm}

\noindent \textit{Techniques:} The key idea of the proof comes from the analysis of random states given in Fig. \ref{fig:hiding}.  There we could find a partition of the system into three regions $ABC$ such that $A$ was approximately decoupled from $B$ (i.e. $\rho_{AB}$ was close to $\rho_A \otimes \rho_B$ in trace norm), and the state on $A$ was close to the maximally mixed state. Then by Uhlmann's theorem \cite{Uhlmann76,SchumacherW-decoupling} it follows that $A$ is strongly correlated with $C$. Therefore the fact that the entropy on $AB$ is close to maximum implies a lower bound on the correlations between $A$ and $C$. In the general case $AB$ might not have entropy close to maximum, and so $A$ might not be decoupled from $B$. However one can still try to follow the same reasoning by applying a measurement on $A$ that \textit{decouples} it from $B$, i.e. makes the post-measurement state on $AB$ close to product.

The above problem -- of decoupling $A$ from $B$ while making the state on $A$ maximally mixed -- was studied before and is known as quantum state merging \cite{HOW05, HOW07}. The name quantum state merging comes from the fact that in decoupling $A$ from $B$ one \textit{merges} the original state in subsystem $A$ to subsystem $C$, with $B$ being the reference subsystem (meaning that the $B$ system holds a purification of the $AC$ system: $\ket{\psi}_{ABC}$). Here we will not be interest in this aspect of the protocol, but merely in the entanglement distillation rate of the process. Given the state $\ket{\psi}_{ABC}^{\otimes n} \in ({\cal H}_A \otimes {\cal H}_B \otimes {\cal H}_C)^{\otimes n}$, with $n$ sufficiently large, consider a random measurement on $\left( {\cal H}_A \right)^{\otimes n}$, consisting of $N \approx 2^{n I(A:B)}$ Haar distributed projectors $\{ P_k \}_{k=1}^N$ of equal dimension summing up to the identity. Here 
\begin{equation}
I(A:B) := H(A) + H(B) - H(AB) 
\end{equation}
is the mutual information of $A$ and $B$. It was shown in Refs. \cite{HOW05, HOW07} that with high probability the post-measurement state $\ket{\phi}_{A'B^nC^n} := (P_k \otimes \id_{B^nC^n})\ket{\psi}_{ABC}^{\otimes n}/\Vert  (P_k \otimes \id_{B^nC^n})\ket{\psi}_{ABC}^{\otimes n} \Vert$ is such that, if $H(B) \leq H(C)$,
\be
\rho_{A'B^n} \approx \tau_{A'} \otimes \rho_{B^n}  \hspace{0.2 cm} \text{and}  \hspace{0.2 cm}  |A'|\approx 2^{-n H(A|C)}, 
\ee
with $\tau_{A'}$ the maximally mixed state on $A'$ and $\rho_{A'B^n}$ the $A'B^n$ reduced density matrix of $\ket{\phi}_{A'B^nC^n}$. We say 
\be
- H(A|C) := H(C) - H(AC) = H(C) - H(B) 
\ee
is the entanglement distillation rate of the protocol, as it gives the number of EPR pairs $\ket{\phi_2} := (\ket{00} + \ket{11})/\sqrt{2}$ shared by $A$ and $C$ after the random measurement on $A$. Here $H(A|C)$ is the conditional entropy of $A$ given the side information $C$. Thus considering many copies of the state and making an appropriate measurement on $A$ we end up again in the situation where $A'$ is close to maximally mixed and decoupled from $B$, implying that if $H(A|C) < 0$, $A'$ is maximally entangled with (part of) $C$. Noting that a maximally entangled state always displays strong correlations, the argument thus suggests that in order not to have long-range correlations between $A$ and $C$ one must have $H(C) \leq H(B)$, which gives an area law for region $C$ if $B$ has constant size.

There are two challenges for turning this idea into a proof. The first concerns the fact that the state merging protocol of \cite{HOW05, HOW07} is devised only in the limit of infinitely many copies of the state, but in our problem we have only a single copy of it. The second is the fact that we only get a particular outcome $k$ only with probability $\approx 2^{- n I(A:B)}$. 

The first challenge is addressed by considering the recent framework of single-shot quantum information theory  \cite{Ren05, Tom12, TCR10, DBWR10, TCR09, JRS07}, which analyses information-theoretical problems in the regime of a single copy of the communication resource (e.g. a quantum state or a quantum channel). For example, one can analyse the rates of state merging when one only has a single copy of the state. For that purpose consider the following analogue of the conditional entropy of a state $\rho_{AB}$, called \textit{min} conditional entropy \cite{TCR10}:
\begin{equation}
H_{\text{min}}(A|B)_{\rho} := \max_{\sigma} \sup \{ \lambda : 2^{- \lambda} I_A \otimes \sigma_B \geq \rho_{AB}    \},
\end{equation}
and the $\varepsilon$-smooth min-entropy of $A$ conditioned on $B$ given by $H_{\text{min}}^{\varepsilon}(A|B)_{\rho} := \max_{\overline{\rho}_{AB} \in {\cal B}_{\varepsilon}(\rho_AB)} H_{\text{min}}(A|B)_{\overline{\rho}}$. Here ${\cal B}_{\varepsilon}(\rho_AB)$ is an $\varepsilon$-ball of states around $\rho_{AB}$ in trace norm. 

Given $\rho_{AC}$ consider an arbitrary purification $\ket{\phi}_{ABC}$ of $\rho_{AC}$. Then the $\varepsilon$-smooth \textit{max} entropy of $A$ conditioned on $C$ is defined by duality as \cite{TCR10}:
\begin{equation}
H_{\text{max}}^{\varepsilon}(A|C) := - H_{\text{min}}^{\varepsilon}(A|B).
\end{equation}

With these analogues of the conditional entropy we are in position of stating the single-shot version of state merging. As shown in Ref. \cite{DBWR10}, there is a protocol that distils a number of EPR pairs (up to error $\varepsilon$ in trace norm) given approximately by $-H^{\varepsilon}_{\text{max}}(A|C)$, using a classical communication cost of approximately $I^{\varepsilon}_{\text{max}}(A:B) := H_{\text{max}}^{\varepsilon}(A) - H_{\text{min}}^{\varepsilon}(A|B)$. Here $H_{\text{max}}(A) := \log \text{rank}(\rho_A)$ is the max entropy of $A$ and $H_{\text{max}}^{\varepsilon}(A) := \min_{\overline{\rho}_{A} \in {\cal B}_{\varepsilon}(\rho_A)} H_{\text{max}}(\overline{\rho}_A)$. Noting that we can write $I(A:B) = H(A) - H(A|B)$, we see that $I^{\varepsilon}_{\text{max}}(A:B)$ plays the role of the mutual information in the single-shot protocol.

The second challenge, in turn, is addressed by applying the merging protocol in \textit{different} partitions concurrently and exploring exponential decay of correlations in each of them. This, together with a result of \cite{Has07} concerning the saturation of mutual information between neighbouring regions in different length scales, is enough to complete the argument. We outline the main steps used in the methods section. A full proof is given in \cite{BH12}.

\vspace{0.3 cm}

\textit{Conclusions:} In this work we proved that for one-dimensional quantum states an area law for their entanglement entropy 
follows merely from the fact that the state has exponential decay of correlations. While intuitively very natural, the relation of exponential 
decay of correlations and area law was put into check by the peculiar kind of correlations embodied in the so-called quantum data hiding
states. The results of the paper thus show that, despite the difficulties caused by these type of correlations, the physically motivated intuition 
is nonetheless correct. In a sense the obstruction provided by ideas from quantum information theory, namely the concept of data hiding states, 
can be overcome by considering the problem also from the perspective of quantum information theory. In particular we employed the central idea 
in quantum Shannon theory of decoupling two quantum systems by performing a random measurement in one of them, as well as recent developments 
in the framework of single-shot quantum information theory. In this respect the results of this paper represent an interesting application of quantum information theory to another area of physics. It is the hope that the approach we developed will lead to further results on the intersection of information theory and many-body physics.



\section{Methods}

Here we give a sketch of proof of Theorem \ref{maintheorempure}, a full proof is presented in \cite{BH12}. First we will present the argument using the simplifying assumption that the asymptotic results (rigorously only valid in the limit of infinitely many copies of the state) hold true for a single copy of the state. Then we present the modifications necessary in the single-shot framework.

\subsection{Ideas of the proof under i.i.d. simplification}

Considering the simplifying assumption that the merging protocol works for a single copy of the state, the upshot of the protocol of \cite{HOW07} is that for a tripartite pure state $\ket{\psi}_{ABC}$, $H(C) \geq H(B)$ implies $\text{Cor}(A:C) \geq 2^{-I(A:B)}$. Indeed, one could first make a random measurement on $A$ obtaining one of the possible outcomes with probability $2^{-I(A:B)}$, distil a maximally entangled state between $A$ and $C$, and then measure the correlations in the maximally entangled state.  We can also write the previous relation as saying that for a pure state $\ket{\psi}_{ABC}$:
\begin{equation} \label{corfromentropybymerging}
\text{Cor}(A:C) \leq 2^{-I(A:B)}  \hspace{0.2 cm} \text{implies} \hspace{0.2 cm} H(C) \leq H(B).
\end{equation}

\begin{figure}[h]
\begin{center}  
\includegraphics[width=9cm,angle=0]{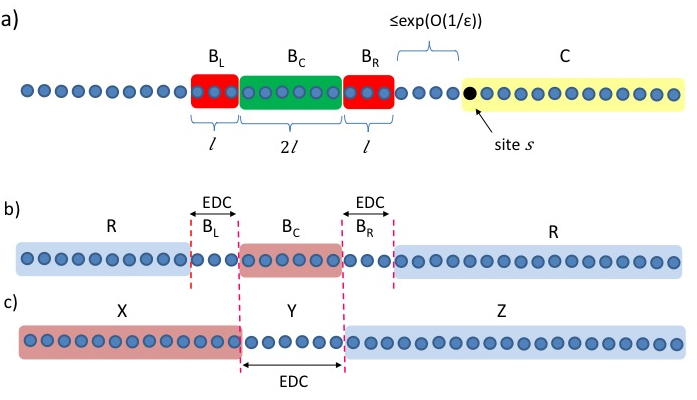}  
\caption{\small \textit{Revealing Correlations: Main Steps of the Proof} (a) Saturation lemma: the mutual information between regions $B_C$ and $B_L B_R$ satisfies $I(B_C:B_LB_R)\leq \epsilon l$; (b) Using state merging to infer subvolume law from saturation of mutual information; (c) Using 
state merging to infer area law from subvolume law. EDC stands for exponential decay of correlations; the pink region plays the role of the party who makes the random measurement in the state merging protocol, the blue region plays the role of the party who obtains the other half of the maximally entangled state, and the white region plays the role of the reference party who does not actively participate in the protocol.
 \label{fig:saturation}}  
\end{center}  
\end{figure}

We split the proof of Theorem \ref{maintheorempure} into three parts. 

\vspace{0.2 cm}

\noindent \textit{1. Area Law from Subvolume Law}: The first step is to show how we can obtain an area law from the assumption that a subvolume law holds true. Consider Fig. \ref{fig:saturation} (c) where the chain is split into three regions $X$, $Y$, and $Z$, with $Y$ composed of $2l$ sites. Suppose that $H(Y) \leq l / \xi$ (the subvolume assumption). Then by EDC (exponential decay of correlations)  we find that 
\begin{equation}
\text{Cor}(X:Z) \leq 2^{- 2l / \xi} \leq  2^{- 2H(Y)} \leq 2^{-I(X:Y)},
\end{equation}
where we used the well-known bound $I(X:Y) \leq 2 H(Y)$ in the last inequality. Therefore by Eq. (\ref{corfromentropybymerging}) we get $H(Z) \leq H(Y)$. 

\vspace{0.2 cm}

\noindent \textit{2. Subvolume Law From Small Mutual Information}: The second step is to show that we can get a region with subvolume law (as required for the first step) from the assumption that there is a region with small mutual information. Consider Fig. \ref{fig:saturation} (b)  with $B_L$ consisting of $2l$ sites, and $B_L$ and $B_R$ consisting of $l$ sites each, and $R$ being the remaining region of the chain. Suppose that $I(B_C : B_L B_R) \leq l / \xi$ (the small mutual information assumption). Then by EDC we have that 
\begin{equation}
\text{Cor}(B_C : R) \leq 2^{-l / \xi} \leq 2^{- I(B_C : B_L B_R) }.
\end{equation} 
Then by Eq. (\ref{corfromentropybymerging}), $H(R) \leq H(B_L B_R)$. Thus 
\begin{eqnarray}
H(B_C) &\leq& H(B_C) + H(B_L B_R) - H(R) \\ &=& I(B_C : BLB_R), \nonumber
\end{eqnarray}
where we used that $H(R) = H(B_C B_L R)$ as the state $\ket{\psi}_{B_L B_C B_R R}$ is pure. But since by assumption $I(B_C : B_L B_R) \leq l / \xi$, we find that indeed $H(B_C) \leq l /\xi$. 

\vspace{0.2 cm}

\noindent \textit{3. Region of Small Mutual Information}: The last part of the proof is to show that there is always a region of small mutual information. Here we use a result concerning the saturation of mutual information in a multiparticle state on different length scales. Consider Fig. \ref{fig:saturation} (a). The result states that for all $\varepsilon > 0$ and a particular site $s$ there exist neighbouring regions $B_{L}B_C B_R$ at a distance at most $\exp(O(1/\varepsilon))$ sites from $s$, with $B_L$ and $B_R$ each of size $l$ and $B_C$ of size $2l$, such that 
\begin{equation}
I(B_C : B_L B_R) \leq \varepsilon l,
\end{equation}
and $l \leq \exp(O(1/\varepsilon))$. In words there is a region $B_LB_CB_R$ of bounded size and distance from $s$ which has sublinear mutual information. This result is a minor adaptation of a result implicit shown by Hastings in \cite{Has07}, and follows easily by successive applications of subadditivity of the von Neumann entropy (see Ref. \cite{BH12} for a proof). 

\vspace{0.2 cm}

We combine the three steps as follows. First we choose the region $C$ for which we would like to prove an area law (see Fig. 3 (a)). Then from the third step we can find regions $B_L, B_C, B_R$, of sizes $l, 2l$, and$ l$, respectively, with $l = \exp(O(\xi))$, and at most $\exp(O(\xi))$ sites away from the boundary of $C$, such that $I(B_C : B_L B_R) \leq l /\xi$. Then from step two we find that $H(B_C) \leq l/\xi$. From step one, in turn, it follows that $H(Z) \leq H(B_C) \leq \exp(O(\xi))$ (where we set $Y = B_C$). Noting that $Z$ and $C$ differ by at most $\exp(O(\xi))$, we find that their entropies can only differ by the same amount. Thus we get that indeed $H(C) \leq \exp(O(\xi))$.

\subsection{Single-shot modifications}

Theorem \ref{maintheorempure} follows from the following area law for the single-shot max entropy:

\def\entropyboundsingle{

Let $\ket{\psi}_{1, ..., n}$ be a state defined on a ring with $\xi$-exponential decay of correlations. Then for any connected region $X \subset [n]$,
\begin{equation} \label{entropyboundsingle}
H^{2^{- \frac{l}{8\xi}}} _{\max}(X) \leq c' \exp\left( c \log(\xi)\xi \right) + l,
\end{equation}
with $c, c' > 0$ universal constants.}

\begin{thm} \label{maintheorempuresingle}
\entropyboundsingle
\end{thm}

It is well known that Theorem \ref{maintheorempuresingle} implies Theorem \ref{maintheorempure} \cite{Has07, VC06}.

We now sketch the proof of Theorem \ref{maintheorempuresingle}, adapting the three steps of the previous section to the proper single-shot framework. In order to do so we use the following two analogues of Eq. (\ref{corfromentropybymerging}):
\begin{equation} \label{statemergingsingleshot1}
- H_{\text{max}}^{o(1)}(A|C) \lesssim 0  \hspace{0.2 cm}  \text{implies}  \hspace{0.2 cm}  \text{Cor}(A:C) \gtrsim 2^{- I_{\max}^{o(1)}(A:B)},  
\end{equation}
and
\begin{equation} \label{statemergingsingleshot2}
H_{\text{max}}^{o(1)}(C) \gtrsim 2 H_{\text{max}}^{o(1)}(B)   \hspace{0.2 cm}  \text{implies}  \hspace{0.2 cm}  \text{Cor}(A:C)  \gtrsim 2^{- 3 H^{o(1)}_{\text{max}}(B)}.
\end{equation}
The notation $- H_{\text{max}}^{o(1)}(A|C) \lesssim 0$ means that $- H_{\text{max}}^{\delta}(A|C) \leq f(\delta)$ for an arbitrary small $\delta > 0$, where $f(\delta)$ is a function only of $\delta$. In this sketch we will assume that the inequalities are perfectly satisfied without the correction terms given by $f(\delta)$. The reader is referred Ref. \cite{BH12} for details of the argument considering all the error terms.

Eq. (\ref{statemergingsingleshot1}) follows directly from the single-shot state merging protocol of \cite{DBWR10}, outlined in the main text. Eq. (\ref{statemergingsingleshot2}), in turn, is a new result proven in \cite{BH12}, using similar techniques to the ones used in the other results. 

\vspace{0.2 cm}

\noindent \textit{1. Area Law from Subvolume Law}: Consider Fig. \ref{fig:saturation} (c) again and suppose now that $H^{o(1)}_{\text{max}}(Y) \lesssim l / \xi$. Then by EDC we find that $\text{Cor}(X:Z) \leq 2^{- 2l / \xi} \lesssim 2^{- H^{o(1)}_{\max}(Y)}$. Then by Eq. (\ref{statemergingsingleshot2}), $H_{\text{max}}^{o(1)}(Z) \lesssim 2 H_{\text{max}}^{o(1)}(Y)$.

\vspace{0.2 cm}

\noindent \textit{2. Subvolume Law From Small Mutual Information}: Consider Fig. \ref{fig:saturation} (b) and suppose that $I_{\max}^{o(1)}(B_C : B_L B_R) \leq l / \xi$. By EDC we have that 
\begin{equation}
\text{Cor}(B_C : R) \leq 2^{-l / \xi} \lesssim 2^{- I^{o(1)}_{\text{max}}(B_C : B_L B_R)}.
\end{equation}
Then by Eq. (\ref{statemergingsingleshot1}), $H_{\text{max}}^{o(1)}(B_C|R) \gtrsim 0$. Thus $H^{o(1)}_{\text{max}}(B_C) \lesssim H^{o(1)}_{\text{max}}(B_C)  + H_{\text{max}}^{\delta}(B_C|R) $. But since by duality $H_{\text{max}}^{o(1)}(B_C|R) = - H_{\text{min}}^{o(1)}(B_C|B_L B_R)$, we get that 
\begin{eqnarray}
H^{o(1)}_{\text{max}}(B_C) &\lesssim& H^{o(1)}_{\text{max}}(B_C) - H_{\text{min}}^{o(1)}(B_C|B_L B_R) \\ &=& I_{\text{max}}^{o(1)}(B_C : B_L B_R). \nonumber
\end{eqnarray}
As by assumption $I_{\max}^{o(1)}(B_C : B_L B_R) \leq l / \xi$, we find that indeed $H^{\delta}_{\text{max}}(B_C) \lesssim l / \xi$.

\vspace{0.2 cm}

\noindent \textit{3. Region of Small Mutual Information}: The last part of the proof is the most technical. It shows a single-shot analogue of the previous result on the saturation of mutual information, but now using $I_{\text{max}}$ instead \textit{and} using the EDC assumption (it is an open question whether the result holds for $I_{\text{max}}$ \textit{without} the EDC assumption). Consider Fig. \ref{fig:saturation} (a). The result states that given a state with EDC, for all $\varepsilon > 0$ and a particular site $s$ there exist neighbouring regions $B_{L}B_C B_R$ at a distance at most $\exp(O(1/\varepsilon)\log(1/\varepsilon))$ sites from $s$, with $B_L$ and $B_R$ each of size $l$ and $B_C$ of size $2l$, such that 
\begin{equation}
I^{o(1)}_{\text{max}}(B_C : B_L B_R) \lesssim \varepsilon l,
\end{equation}
and $l \leq \exp(O(1/\varepsilon \log(1/\varepsilon)))$. The proof is more involved than in the von Neumann case and uses several techniques of single-shot information theory, such as the quantum equipartition property \cite{TCR09}, and the quantum substate theorem \cite{JRS07}. 

To summarize, we employ the state merging protocol in the form of Eq. (\ref{corfromentropybymerging}) and the assumption of exponential decay of correlations \textit{three} times. One in conjunction with the result about saturation of mutual information in order to get a region of constant size and not so large entropy, a second to boost this into an area law for regions of arbitrary size, and a third to prove the saturation of the single-shot mutual information. This finishes the sketch of the proof. The full proof is given in \cite{BH12}.

{\it Acknowledgement}
We would like to thank Dorit Aharonov, Itai Arad, and Aram Harrow for interesting discussions
on area laws and related subjects and Matt Hastings for useful correspondence. FB acknowledges
support from the Swiss National Science Foundation, via the National Centre of Competence
in Research QSIT. MH thanks the support of EC IP QESSENCE, ERC QOLAPS, and National
Science Centre, grant no. DEC-2011/02/A/ST2/00305. Part of this work was done at National
Quantum Information Centre of Gdansk. F.B. and M.H. thank the hospitality of Institute Mittag
Leer within the program Quantum Information Science (2010), where part of this work was done.

{\it Authors Contribution}
All authors contributed to all aspects of this work

\end{document}